\documentclass{article}

\usepackage[numbers]{natbib}         
\usepackage[colorlinks]{hyperref}    
\usepackage[english]{babel} 
\usepackage{amssymb}
\usepackage{amsmath}
\usepackage{txfonts}
\usepackage{mathdots}
\usepackage[classicReIm]{kpfonts}
\usepackage[dvips]{graphicx} 
\usepackage[a4paper, portrait, margin=1in]{geometry}
\usepackage{tabularx}
\usepackage{longtable}
\usepackage{multirow}
\usepackage{booktabs}
\usepackage[labelsep=period]{caption}
\usepackage{makecell}
\begin{document}
	\setlength{\parindent}{0pt}
	\setlength{\parskip}{1ex}
	
	\textbf{\Large Artificial Intelligence in Tumor Subregion Analysis Based on
		Medical Imaging: A Review}
	
	\bigbreak

	Mingquan Lin, Jacob Wynne, Yang Lei, Tonghe Wang, Walter J. Curran, Tian Liu and Xiaofeng Yang*
	
	Department of Radiation Oncology and Winship Cancer Institute, Emory University, Atlanta, GA 30322

	\bigbreak
	\bigbreak
	\bigbreak

	\textbf{*Corresponding author: }
	
	Xiaofeng Yang, PhD
	
	Department of Radiation Oncology
	
	Emory University School of Medicine
	
	1365 Clifton Road NE
	
	Atlanta, GA 30322
	
	E-mail: xiaofeng.yang@emory.edu

	\bigbreak
	\bigbreak
	\bigbreak
	\bigbreak
	\bigbreak
	\bigbreak

	\textbf{Abstract}

	Medical imaging is widely used in cancer diagnosis and treatment, and artificial intelligence (AI) has achieved tremendous success in various tasks of medical image analysis. This paper reviews AI-based tumor subregion analysis in medical imaging. We summarize the latest AI-based methods for tumor subregion analysis and their applications. Specifically, we categorize the AI-based methods by training strategy: supervised and unsupervised. A detailed review of each category is presented, highlighting important contributions and achievements. Specific challenges and potential AI applications in tumor subregion analysis are discussed. 
	
	\bigbreak
	\bigbreak
	
	\textbf{keywords:} Artificail intelligence (AI), Machine learning (ML), Deep learning (DL), Tumor subregion, Medical imaging

	\noindent 
	\section{ INTRODUCTION}
	
	In current clinical practice and research, tumor is usually assumed to be homogeneous or heterogeneous with similar distribution throughout the entire volume \cite{RN9, RN10, RN12, RN11, RN13}. Recent studies have shown that some tumor regions may be more biologically aggressive than others and may play a dominant role in disease progression \cite{RN15, RN8, RN14}. Neglecting such tumor heterogeneity at various spatial and temporal scales can lead to failures in prognosis and treatment \cite{RN8}. Medical imaging has been shown to be able to reveal and quantify the heterogeneity within tumors \cite{RN189, RN173, RN172}. Individual tumor can then be divided into sub-regions based on detected regional variations. Diagnosis, prognosis, and evaluation of treatment response can be performed individually in these subregions, and has proved superior to a simple analysis of the whole tumor \cite{RN16, RN17}. Therefore, accurate detection and analysis of tumor sub-regions is of great clinical and research interest. 
	
	Over the last few years, artificial intelligence (AI)  has achieved tremendous success in various tasks in the field of medical imaging \cite{RN19, RN23, RN22, RN21, RN24, RN25, RN26,RN20, RN18}. Many AI-based methods have been proposed to locate and analyse tumor subregions for a variety of imaging modalities and clinical tasks. In this study, we review the applications of supervised and unsupervised AI models in imaging-based tumor subregion analysis. With this survey, we aim to: 
	
	1. Summarize the lastest developments of AI applications in imaging-based tumor subregion analysis.
	2. Highlight contributions, identify challenges, and outline future trends.

	\noindent 
	\section{ARTIFICIAL INTELLIGENCE}

	AI is a field that seeks to enable machines to learn from experience,  think like humans, and perform human-like tasks. Machine learning (ML) is a discipline within AI,  in which computers are trained to automatically improve performance on specific tasks based on experience. Training methods in ML are broadly composed of supervised, semi-supervised, or unsupervised strategies, each with decreasing need for human input. Within ML, deep learning (DL) employs multi-layer (“deep”) networks of mathematical functions initially intended to imitate the structure and function of the human brain to fundamentally create a mapping from one representational domain to another (e.g. categorizing photos to names of the objects they contain). Both supervised and unsupervised methods are commonly used in DL for medical image analysis. 
	
	\noindent 
	\subsection{Supervised learning}
	In supervised learning, an algorithm is designed to learn a mapping function $f(\bullet)$ from the input variable (x) to the output variable (Y), i.e. Y=f(x). The goal is to approximate the mapping function well so that the output variable (Y) of new input data (x) can be accurately predicted.  Least-absolute-shrinkage-and-selection-operator (Lasso), random forest (RF), support vector machine (SVM), and artificial neural networks (ANN) are widely used algorithms in determing the mapping function. The Lasso is a shrinkage and feature selection method for linear regression  \cite{RN27}. It minimizes the sum of squared errors and the sum of the absolute value of coefficients.  RF is an ensemble learning algorithm that boosts performance by combining the results of many weaker algorithms effectively reducing overfitting and building a model that is robust for discrete values in the feature space  \cite{RN28}. The object of SVM is to find a hyperplane in n-dimensional space that maximizes the separation of different classes of data in the feature space \cite{RN29}. 
	
	The multilayer perceptron (MLP) is a class of feedforward ANN wherein the biological unit of the brain ,the neuron, is modeled by the mathematical unit of a network node \cite{RN30}. An MLP consists of at least three layers of nodes: an input layer, hidden layer, and output layer. All nodes except the inputs employ nonlinear activation functions. MLP uses a supervised learning technique called backpropagation to update the parameters of each node. The multilayer structure and nonlinear activation of MLP distinguish it from linear perceptrons and allow it to distinguish data that are not linearly separable. Although MLP has been successfully applied to practical problems in many fields, these models must be carefully trained and thoughtfully deployed to avoid overfitting or, alternatively, failure of convergence during inference.
	
	Convolutional neural networks (CNN) have been widely applied in many tasks \cite{RN49, RN185, RN186, RN48, RN177, RN176, RN187, RN183}. A typical CNN may be composed of several layers performing discrete computational tasks including: convolution at various scales of resolution, maximum or other forms of pooling, and batch normalization. The outputs of these layers may be omitted as in dropout or be passed as inputs to all subsequent layers when fully connected layers are employed. In order to improve the performance of  deep CNNs, various architectures have been proposed. U-Net adopts symmetrical encoding and decoding paths with skip connections between them and is widely used in medical image segmentation. The residual network (ResNet) architecture employs a shortcut connection which reduces the likelihood of “vanishing” gradients during training, allowing the development of deeper networks.
	
	\noindent 
	\subsection{Unsupervised learning}
	Supervised learning requires time-consuming and labor-intensive manual data annotations. In contrast, unsupervised techniques learn the distribution of input data and divide samples into clusters without labeled training dataset. Common unsupervised learning algorithms include the active contour model (ACM), hidden markov random fields (HMRF), the K-means and expectation-maximization (EM) algorithms, principal component analysis (PCA) and hybrid hierarchical clustering. ACM works to segment objects in an image by evolving a curve according to the constraints in the image \cite{RN32}. The HMRF model is a random process generated by MRF. Its state sequence cannot be directly observed, but can be indirectly estimated through observation \cite{RN33}. The EM algorithm is an iterative method that searches the (local) maximum likelihood or maximum a posteriori (MAP) estimate of the parameters in a statistical model \cite{RN34}. PCA is an orthogonal linear transformation that reduces the dimensionality of the input data while retaining its most significant parts \cite{RN35}. K-means identifies k centroids and assigns each data point to the nearest centroid by minimizing the sum of the squared Euclidean distances between each point and its assigned centroid \cite{RN36}.  Hybrid hierarchical cluster combines the advantages of bottom-up hierarchical clustering and top-down clustering,  so it is applicable to various sizes of data  \cite{RN37}.

	\noindent 
	\section{LITERATURE SEARCH}
	The scope of this review is limited to the applications of AI in tumor subregion analysis. Peer-reviewed journal publications appearing after December 31, 2016 were collected from various databases including Google Scholar, PubMed, Web of Science, etc. We used a variety of keywords such as machine learning, deep learning, intratumor, subregion, subvolume, voxel-based, overall survival, clustering. Publications describing the methods of the top three performers in the Brain tumor segmentation (BraTS) challenge from 2017-2019 were included. For all other body sites, the included publications are listed in tables accompanying each dedicated section. A total of 89 papers were identified discussing AI applications in imaging-based tumor subregion analysis. The number of publications is plotted by year in Fig 1.
	
	\begin{figure}
		\centering
		\noindent \includegraphics*[width=6.50in, height=4.20in, keepaspectratio=true]{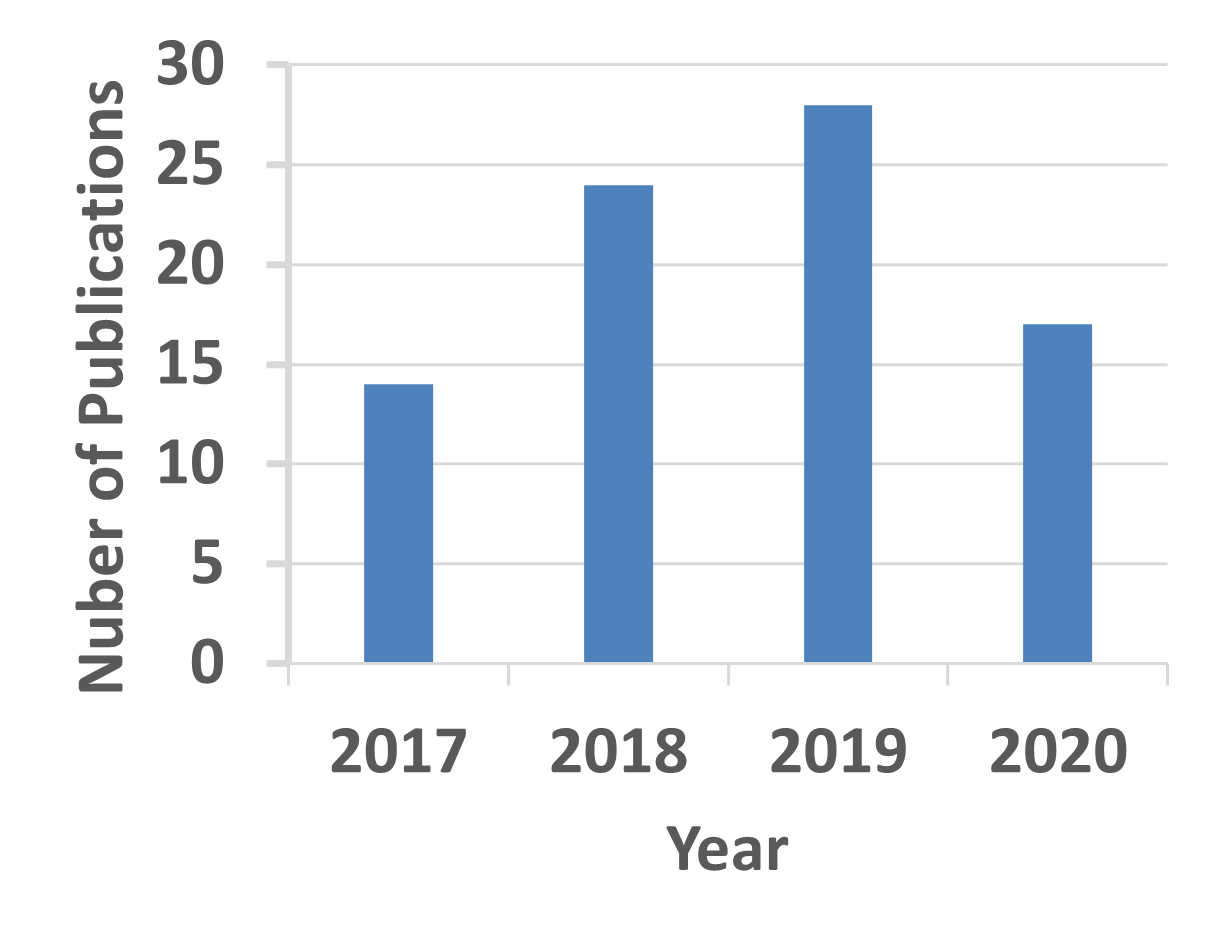}
		
		\noindent Fig. 1. Number of publications in AI-based tumor subregion analysis. “2020” only covers the first five months of 2020. 
	\end{figure}

	\noindent 
	\section{AI IN TUMOR SUBREGION ANALYSIS OF MEDICAL IMAGES}
	
	\noindent 
	\subsection{Supervised learning in tumor subregion analysis of medical images}
	Supervised learning has been widely used in tumor subregion analysis for identification of recurrence volume, prediction of outcomes including overall survival (OS) or progression-free survival (PFS), and subregion segmentation.  Sixty-four papers related to supervised learning are included in this paper.
	
	\noindent
	\subsubsection{Head and neck (HN)}
	CT and 18-FDG PET are often used in staging, radiation therapy treatment planning and evaluation of treatment response in patients with cancers of the head and neck \cite{RN174, RN175}. PET provides detailed functional and metabolic molecular information, while CT reveals the precise anatomical position of the tumor. Table 1 shows a list of selected studies that used supervised learning in tumor subregion analysis based on medical images in the head and neck. Ding et al. investigated the clinicopathological characteristics of different supraglottic subregions and their correlation with the prognosis of patients with squamous cell carcinoma \cite{RN42}. Supraglottic squamous cell carcinomas were divided into four types based on subregion: epiglottis, ventricular bands, aryepiglottic fold, and ventricle. A Cox proportional hazards model was used to generate a biomarker. They found that there were significant differences in the regional control rate, overall survival rate, and cancer-specific survival rates among different subregions, indicating that patients with carcinoma of the epiglottis or ventricular bands had an increased survival rate relative to those with disease in the aryepiglottic fold or ventricle. Beaumont et al. \cite{RN43} developed a voxel-wise ML model to identify the sub-regions with tumor recurrence and to predict their location based on pre-treatment PET images. A RF model was trained with voxel-wise features. Voxel-wise analysis based on radiomic features and spatial location within the tumor was shown helpful in determining the location of recurrence and providing guidance to tailor chemoradiation therapy (CRT) through dose escalation within the area of radiation resistance.
	
\begin{table}[]
		\caption{Overview of supervised learning for tumor subregions analysis based on medical imaging for HN.}
	\begin{tabular}{llllll}
		\hline
		Ref                          & Year & Model                                                                     & Task                                                                        & Modality       & \begin{tabular}[c]{@{}l@{}}\# of patients in \\ training/testing datasets\end{tabular} \\ \hline
		\cite{RN42} & 2017 & \begin{tabular}[c]{@{}l@{}}Cox proportional \\ hazards model\end{tabular} & Predict OS                                                                  & MRI, CT        & 111 (N/A)                                                                              \\
		\cite{RN43} & 2019 & RF                                                                        & \begin{tabular}[c]{@{}l@{}}Recurrence volume \\ identification\end{tabular} & 18F-FDG PET/CT & 26, LOOCV                                                                              \\ \hline
	\end{tabular}

LOOCV: leave-one-out cross-validation. N/A:not available,  indicating that the paper only provides the total number of samples.
\end{table}

\noindent
\subsubsection{Gliomas}
	
	Gliomas are the most common primary brain tumor and can be classified by histopathologic features into two groups: high-grade gliomas (HGG) and low-grade gliomas (LGG). Magnetic resonance imaging (MRI) is the main imaging modality to noninvasively diagnose brain tumors by providing high soft tissue contrast \cite{RN44}.  Dividing gliomas into substructures played an important role in glioma diagnosis, staging, monitoring and treatment planning for patients. Table 2 shows a list of selected studies using supervised learning in tumor subregion analysis based on medical images for gliomas. 
	
	Fiouznia et al. \cite{RN45} developed a model to discriminate glioma tissue subregions based on multiparametric (mp) MRI. Based on the histopathological results,  subregions were categorized into active tumor (AT), infiltrative edema (IE), and normal tissue (NT). In the study of Fischer et al., linear discriminant analysis (LDA), quadratic discriminant analysis (QDA), and SVM were applied to distinguish the three tissue subtypes from each other based on selected features derived from sub-regions. All three classifiers achieved the high classification performance (AUC ~ 90\%) with a combination of “CBV, MD, $T_2$\_ISO, FLAIR” features . This capability might be advantageously employed to locate tissue subregions prior to image-guided biopsy procedures. Some studies further predicted OS or PFS based on tumor subregion analysis  \cite{RN47, RN46}. 
	
	Zhou et al. \cite{RN46} developed a framework to identify tumor subregions based on pretreatment MRI for patients with glioblastoma (GBM), correlating the image-based spatial characteristics of subregions with survival rate. Two datasets were included in this study. The habitat-based features were extracted from the GBM subregions derived from intratumoral grouping and spatial mapping. The results revealed that habitat-based features were effective for predicting two survival groups (accuracy is 87.5\% and 86.36\%, respectively). The results generated by classifiers (SVM, k-nearest neighbors (KNN), and na$\ddot{i}$ve Bayes) showed that the spatial correlation features between the signal-enhanced subregions can effectively predict survival group (P < 0.05 for all classifiers).  GBM is further characterized by infiltrative growth at the cellular level that cannot be completely resected. Diffusion tensor imaging (DTI) has been shown to potentially detect tumor infiltration by reflecting microstructural destruction. To investigate the incremental prognostic value of infiltrative patterns over clinical factors and identify specific subregions that may be suitable for targeted therapy, Li et al. \cite{RN47} explored the heterogeneity of GBM infiltration using joint histogram analysis in DTI. The prognostic value of covariates for OS and PFS at 12 and 18 months was tested using a logistic regression model. The results showed that joint histogram features have incremental prognostic values when combined with clinical factors, suggesting that patients may benefit from adaptive radiation therapy strategies based on prognostic data obtained during and after treatment if these high-risk tumor subregions can be identified.
	
	CNN have achieved tremendous success in  tumor subregion analysis and can used to extract features and segment tumor subregions. Small sample size is one problem encountered with the application of CNNs to limited medical images. Transfer learning and fine-tuning may be employed to ameliorate small sample problems, making CNN more applicable to various medical image tasks \cite{RN50}. Lao et al. extracted features from manually segmented tumor subregions based on multi-modality MR images and used these features to generate a proposed signature based on LASSO \cite{RN51}. The extracted features included two parts: hand-crafted features and features extracted by a pre-trained deep DL model. The study demonstrated that transfer learning-based deep features were able to generate imaging signatures for OS prediction and risk stratification for GBM, indicating the potential of DL feature-based biomarkers in the preoperative care of patients with GBM.  CNNs can also be used to segment tumor subregions to facilitate their further study. Based on multiparametric MRI, Kazerooni et al. \cite{RN52} constructed a multi-institutional radomics model that supports the upfront prediction of PFS and recurrence pattern (RP) in patients diagnosed with GBM at the time of initial diagnosis. The proposed framework included subregion identification (DeepMedic \cite{RN53}), feature extraction, sequential forward feature selection, biomarker generation, and classification using a SVM implemented using the Cancer Imaging Phenomics Toolkit (CaPTk) open-source software.  The area under the operator-received curve (AUC) for PFS prediction was 0.88 and 0.82-0.83; AUC for RP was 0.88 and 0.56-0.71 for the single-institution and multi-institutional analyses, respectively. The results suggest that the biomarkers included in the radiomics models as implemented in CaPTK could predict PFS and RP in patients diagnosed with GBM.  Isocitrate dehydrogenase 1 (IDH1) is established as a prognostic and predictive marker for patients with GBM \cite{RN56,RN60, RN57, RN55, RN58, RN59}. Li et al. \cite{RN54} developed a model to predict IDH mutation status in GBM preoperatively based on multiregional radiomic features derived from mpMRI. The proposed model was tested on an independent validation cohort. IDH1 mutation was predicted by the RF model after using Boruta \cite{RN61} for feature selection. The multi-tumor subregions were automatically segmented using a CNN \cite{RN62}. The model’s best performance achieved 97\% accuracy with AUC 0.96, and F1‐score 0.84. The multi-region model built using all-region features performed better than single-region models. The multi-region model achieved the best performance when combining age with all-region features. The results showed that the proposed model based on multi-regional mpMRI features has the potential to detect IDH1 mutation status in GBM patients prior to surgery.
	
	\begin{table}[]
		\caption{Overview of supervised learning for tumor subregions analysis based on medical imaging for gliomas.}
		\begin{tabular}{lllllll}
			\cline{1-6}
			Ref                           & Year & Models                                                                    & Task                                                                                           & Modality                                                                                               & \begin{tabular}[c]{@{}l@{}}\# of patients in \\ training/testing datasets\end{tabular} &  \\ \cline{1-6}
			\cite{RN45}  & 2018 & LDA, QDA, SVM                                                             & \begin{tabular}[c]{@{}l@{}}Predict active and \\ infiltrative tumorous \\subregions\end{tabular} & \begin{tabular}[c]{@{}l@{}}T1W, T2W, \\FLAIR, \\T2-relaxometry,  DWI, \\DTI, IVIM, and DS-MRI\end{tabular} & 10, LOOCV                                                                              &  \\
			\cite{RN46}  & 2017 & \begin{tabular}[c]{@{}l@{}} SVM, KNN,\\ Nave  Bayes   \end{tabular}                                                   & Predict overall survival                                                                       & T1W-ce, FLAIR, T2W                                                                                     & 79, LOOCV                                                                              &  \\
			\cite{RN47}  & 2019 & logistic regression                                                       & \begin{tabular}[c]{@{}l@{}}Identify specific subregions \\ for targeted therapy\end{tabular}   & DTI                                                                                                    & 115,(N/A*)                                                                             &  \\
			\cite{RN51}  & 2017 & CNN, LASSO                                                                & Predict OS                                                                                     & T1W, T1-Gd, FLAIR, T2W                                                                                 & 75/37                                                                                  &  \\
			\cite{RN52}  & 2020 & DeepMedic, SVM                                                            & Predict PFS and RP                                                                             & \begin{tabular}[c]{@{}l@{}}T1W, T1-Gd, FLAIR, \\ T2W, DWI, DS-MRI\end{tabular}                         & \begin{tabular}[c]{@{}l@{}}Scheme 1 and 3:80, 10 fold \\ Scheme 2: 56/24\end{tabular}  &  \\
			\cite{RN54}  & 2018 & RF                                                                        & \begin{tabular}[c]{@{}l@{}}Predict  isocitrate \\ dehydrogenase \\1 genes (IDH1)\end{tabular}    & T1W, T1-Gd, FLAIR, T2W                                                                                 & 118/107                                                                                &  \\
			\cite{RN144} & 2018 & RF                                                                        & Predict survival time                                                                          & T1W-ce, FLAIR                                                                                          & 73, LOOCV                                                                              &  \\
			\cite{RN145} & 2018 & RF                                                                        & Predict OS and PFS                                                                             & T1W, FLAIR                                                                                             & 40, 5 folds                                                                            &  \\
			\cite{RN146} & 2019 & SVM                                                                       & Glioma grading                                                                                 & \begin{tabular}[c]{@{}l@{}}DTI, T1W-ce, \\FLAIR,  T2W-FSE,\\ DSCE-RAW, 1H-MRS\end{tabular}               & 40, LOOCV                                                                              &  \\
			\cite{RN147} & 2019 & LASSO                                                                     & \begin{tabular}[c]{@{}l@{}}stratify glioblastoma \\ patients basd on survival\end{tabular}     & \begin{tabular}[c]{@{}l@{}}T1W, T1W-CE, \\ FLAIR, T2W\end{tabular}                                     & 70/35                                                                                  &  \\
			\cite{RN148} & 2019 & \begin{tabular}[c]{@{}l@{}}Cox proportional \\ hazards model\end{tabular} & \begin{tabular}[c]{@{}l@{}}stratify glioblastoma \\ patients basd on survival\end{tabular}     & post-T1W                                                                                               & 85/42                                                                                  &  \\
			\cite{RN167} & 2018 & CNN                                                                       & \begin{tabular}[c]{@{}l@{}} Tumor subregions \\segmentation            \end{tabular}                                                     & T1W-CE                                                                                                 & 186/47                                                                                 &  \\ \cline{1-6}
		\end{tabular}
	
	* Exact training and testing datasets are not available.
	\end{table}
	
	\noindent 
	\subsubsection{BraTS challenge}
	
	As mentioned above, glioma subregion segmentation may play an important role in future glioma diagnosis, staging and treatment planning. Most of the research described here uses a non-public or institutional dataset, making it difficult to compare methods or results against other published work. The BraTS challenge stands in contrast to these, providing pre-operative mpMRI scans sourced from multiple institutions to inspire and evaluate the reproducibility of state-of-the-art methods for glioma brain tumor segmentation \cite{RN66, RN65, RN64, RN67, RN63}. The data set includes images of four MR sequences: T1, T1-Gd, T2, and FLAIR. The labels are divided into four classes (0: healthy tissues, 1: necrosis and non-enhancing tumor, 2: edema, 4: enhancing tumor). The evaluation system divides the tumor into three regions for performance evaluation according to practical clinical application: (1) the whole tumor (WT) region with labels 1, 2, and 4; (2) the tumor core (TC) with labels 1 and 4; (3) the enhancing tumor (ET) region (lable 4).
	
	Table 3 contains a list of selected references using supervised learning in tumor subregion analysis of BraTS challenge data. Most of these are based on DL with various proposed architectures. Attention gates are commonly adopted to improve performance due to its utility in automatically highlighting informative elements of intermediate feature maps. Hu et al. proposed a novel 3D refinement module that can aggregate local detail information and 3D semantic context directly within the 3D convolutional layer \cite{RN70}. Kamnitsas et al. developed a 3D-CNN with a dual pathway and 11 convolutional layers \cite{RN53}. In order to cope with the computational burden of the 3D network, the processing of adjacent image paths was combined into a channel through the network during training, while automatically adapting to the inherent class imbalances existing in the data. They used a dual-path architecture to simultaneously process multi-scale input images to obtain multi-scale context information. A 3D fully conditional random field (CRF) was employed in post-processing and was shown to be effective in mitigating false positives. Havaei et al. developed a novel CNN with a two-pathway architecture which was adopted to simultaneously extract both local and global contextual features \cite{RN68}. They modeled local label dependencies by cascade-CNN rather than CRF. This method can significantly improve computational speed by employing the efficient convolution operation rather than CRFs. Due to the tremendous success of the attention mechanism in computer vision at large \cite{RN74, RN73, RN71, RN72, RN76, RN75} and medical image analysis specifically \cite{RN80, RN79, RN82, RN78, RN77, RN95}, Zhang et al. integrated an attention gate into U-Net to generate an Attention Gate Residual U-Net (AGResU-Net) model for brain tumor segmentation \cite{RN83}. Several attention gate units were added to the skip connection of U-Net to highlight contrast information while disambiguating irrelevant and noisy feature responses.
	
	\begin{table}[]
		\caption{Overview of supervised learning in tumor subregion analysis of BraTS challenge data.}
		\begin{tabular}{lllll}
			\hline
			Ref                           & Year & Models                                                                                      & Task                                                                               & \begin{tabular}[c]{@{}l@{}}\# of patients  in \\ training/testing datasets\end{tabular} \\ \hline
			\cite{RN68}  & 2017 & Cascade CNN                                                                                 & Tumor subregion segmnetation                                                       & 60, 7 fold                                                                              \\
			\cite{RN53}  & 2017 & \begin{tabular}[c]{@{}l@{}}Efficient Multi-scale \\ U-Net with CRFs\end{tabular}            & Tumor subregion segmnetation                                                       & 253, 5 fold                                                                             \\
			\cite{RN70}  & 2020 & \begin{tabular}[c]{@{}l@{}}3D refinement\\    \\ U-Net\end{tabular}                         & Tumor subregion segmnetation                                                       & 274/110                                                                                 \\
			\cite{RN83}  & 2020 & Attention Gate ResU-Net                                                                     & Tumor subregion segmnetation                                                       & \begin{tabular}[c]{@{}l@{}}285/46, 285/66,\\ 335/125\end{tabular}                       \\
			\cite{RN149} & 2018 & Emsemble CNN                                                                                & Tumor subregion segmnetation                                                       & 285 (N/A)                                                                               \\
			\cite{RN150} & 2019 & multi-cascaded CNN with CRFs                                                                & Tumor subregion segmnetation                                                       & 40, 274, 285                                                                            \\
			\cite{RN151} & 2019 & 3D dilated multi-fiber U-Net                                                                & Tumor subregion segmnetation                                                       & 285/66                                                                                  \\
			\cite{RN152} & 2020 & Cross-task Guided Attention U-Net                                                           & Tumor subregion segmnetation                                                       & \begin{tabular}[c]{@{}l@{}}274/110, 285/46,\\ 285/66\end{tabular}                       \\
			\cite{RN153} & 2019 & 2D-3D context U-Net                                                                         & Tumor subregion segmnetation                                                       & 235/50/46                                                                               \\
			\cite{RN154} & 2018 & CNN                                                                                         & Tumor subregion segmnetation                                                       & 240/34                                                                                  \\
			\cite{RN155} & 2019 & Inception-based U-Net                                                                       & Tumor subregion segmnetation                                                       & \begin{tabular}[c]{@{}l@{}}165/55/54,\\ 171/57/57\end{tabular}                          \\
			\cite{RN156} & 2018 & FCNN with CRFs                                                                              & Tumor subregion segmnetation                                                       & \begin{tabular}[c]{@{}l@{}}30/35, 274/110,\\ 274/191\end{tabular}                       \\
			\cite{RN157} & 2018 & \begin{tabular}[c]{@{}l@{}}SVM, RF, \\ Logistic regression\end{tabular}                     & Glioma grading                                                                     & 285, 5 fold                                                                             \\
			\cite{RN158} & 2020 & U-Net, RF                                                                                   & \begin{tabular}[c]{@{}l@{}}Tumor subregion segmnetation,\\ Predict OS\end{tabular} & 268/67, 76/29                                                                           \\
			\cite{RN159} & 2019 & LASSO                                                                                       & Predict OS                                                                         & 163, 5 fold                                                                             \\
			\cite{RN168} & 2020 & \begin{tabular}[c]{@{}l@{}}Heterogeneous CNN with \\ CRFs-Recurrent Regression\end{tabular} & Tumor subregion segmnetation                                                       & 60 (N/A*)                                                                               \\
			\cite{RN169} & 2019 & 2.5D cascade CNN                                                                            & Tumor subregion segmnetation                                                       & \begin{tabular}[c]{@{}l@{}}285/46/146,\\ 285/66/191\end{tabular}                        \\
			\cite{RN170} & 2020 & IOU 3D symmetric  fully CNN                                                                 & Tumor subregion segmnetation                                                       & 134/33                                                                                  \\
			\cite{RN171} & 2020 & CNN                                                                                         & Tumor subregion segmnetation                                                       & \begin{tabular}[c]{@{}l@{}}20/10, 192/82,\\ 285/146, 285/191\end{tabular}               \\
			\cite{RN224} & 2020 & CNN, SVM                                                                                    & Tumor subregion segmnetation                                                       & 274, 10 fold                                                                            \\
			\cite{RN225} & 2018 & CNN                                                                                         & Tumor subregion segmnetation                                                       & 274/110                                                                                 \\
			\cite{RN227} & 2019 & CNN                                                                                         & Tumor subregion segmnetation                                                       & 285/46, 285/66                                                                          \\
			\cite{RN228} & 2020 & U-Net                                                                                       & Tumor subregion segmnetation                                                       & 285/46, 285/66                                                                          \\
			\cite{RN229} & 2018 & Hybird pyramid U-Net                                                                        & Tumor subregion segmnetation                                                       & 285, 5 fold                                                                             \\
			\cite{RN230} & 2019 & CNN                                                                                         & Tumor subregion segmnetation                                                       & 285 (N/A)                                                                               \\
			\cite{RN231} & 2020 & CNN                                                                                         & Tumor subregion segmnetation                                                       & 27/254,285                                                                              \\
			\cite{RN232} & 2020 & CNN                                                                                         & Tumor subregion segmnetation                                                       & 85/200                                                                                  \\
			\cite{RN233} & 2020 & CNN                                                                                         & Tumor subregion segmnetation                                                       & 68/8, 50/6                                                                              \\ \hline
		\end{tabular}
	
	* Exact training and testing datasets are not available
	\end{table}

	Table 4 lists the three top-performing studies from 2017 to 2019 with their results. Ensemble learning, cascade learning, and multi-scale operations are commonly added to CNNs to improve the accuracy of  brain tumor subregion segmentation.  In statistics and machine learning, ensemble learning combines models to surpass the performance of any one consitituent model and is commonly used to improve classification, prediction and segmentation performance. Kamnitsas et al. \cite{RN84} developed a framework (EMMA) to combine several DL models for robust segmentation. EMMA independently trained DeepMedic \cite{RN53}, FCN \cite{RN85}, and U-Net \cite{RN3828}, combining their segmentation predictions at testing. Myronenko et al. proposed a semantic segmentation CNN with asymmetric large encoders to segment tumor subregions \cite{RN92}. A variational autoencoder (VAE) branch was added to the network to reconstruct the input images jointly with the segmentation and regularize the shared encoder. Finally, they assembled ten models trained from scratch to further improve performance. Zhao et al. \cite{RN97} developed a self-ensemble U-Net, combining multi-scale prediction to boost accuracy with a slight increase in memory consumption. They also used the average of all models in the final ensemble and averaged the prediction of the overlapping patches to obtain a more accurate result.  Cascade learning is a particular case of ensemble learning based on the concatenation-in-series of several models, using preceding model outputs as inputs for the next model in the cascade. Wang et al. trained three networks for cascade learning, each with a similar structure, including a large encoder part with dilated convolutions and a basic decoder \cite{RN90}. The WT was segmented first and bounding box of the result was used for the TC segmentation. Finally, the ET segmentation was based on the bounding box of the TC segmentation. The 3 × 3 × 3 convolution kernel was decomposed into 3 × 3 × 1 and 1 × 1 × 3 kernels to reduce the number of parameters and deal with anisotropic receptive fields. Jiang et al. \cite{RN96}  developed a two-stage cascaded U-Net to segment brain tumor subregions from coarse to fine-scale. In the first stage, a U-Net predicts a coarse segmentation result based on the multi-modal MRI. The coarse segmentation provides the rough locations of tumors and this is used to highlight contrast information. The coarse segmentation results are combined with the raw input images prior to input into a second U-Net with two decoder paths (one using a deconvolution, the other using trilinear interpolation) to generate a fine segmentation map. Zhou et al. \cite{RN95} proposed an ensemble framework combining different networks to segment tumor subregions with more robust results. The proposed framework considered multi-scale information by segmenting three tumor subregions in cascade with a shared backbone weight and an attention block. Multi-scale and deeper networks may achieve better segmentation results because brain tumors have a highly heterogeneous appearance on MR images. Mckinly et al. \cite{RN94} proposed a U-Net-like network containing a DenseNet with dilated convolutions. The author also introduced a new loss function, a generalization of binary cross-entropy, to solve label uncertainty. In another study, Mckinly et al. \cite{RN98}  used a structure very similar to the previous one,\cite{RN94} but replaced Batch normalization with instance normalization and added a simple local attention mechanism between dilated dense blocks. This study also included more data for training so that it may also improved the performance of the network. Isensee et al. made a minor modifications to U-Net, replacing ReLU and batch normalization with leaky ReLU and instance normalization to achieve competitive performance.\cite{RN93} They also supplemented with data from their own institution to achieve a 2\% increase in Dice similarity coefficient (DSC) on the enhancing tumor training data.
	
	\begin{table}[]
		\caption{Overview of the top 3 segmentation peformance of the last three BraTS (2017-2019).}
		\begin{tabular}{ccccccccc}
			\hline
			\multirow{2}{*}{Ref}         & \multirow{2}{*}{Year} & \multirow{2}{*}{Ranking} & \multicolumn{3}{c}{DSC} & \multicolumn{3}{c}{Hausdorff95 (mm)} \\
			&                       &                          & WT     & TC     & ET    & WT         & TC         & ET         \\ \hline
			\cite{RN84} & 2017                  & 1                        & 0.886  & 0.785  & 0.729 & 5.01       & 23.10      & 36.00      \\
			\cite{RN90} & 2017                  & 2                        & 0.874  & 0.775  & 0.783 & 6.55       & 27.05      & 15.90      \\
			\cite{RN88} & 2017                  & 3                        & 0.858  & 0.775  & 0.647 & N/A        & N/A        & N/A        \\
			\cite{RN91} & 2017                  & 3                        & N/A    & N/A    & N/A   & N/A        & N/A        & N/A        \\
			\cite{RN92} & 2018                  & 1                        & 0.884  & 0.815  & 0.766 & 3.77       & 4.81       & 3.77       \\
			\cite{RN93} & 2018                  & 2                        & 0.878  & 0.806  & 0.779 & 6.03       & 5.08       & 2.90       \\
			\cite{RN94} & 2018                  & 3                        & 0.886  & 0.799  & 0.732 & 5.52       & 5.53       & 3.48       \\
			\cite{RN95} & 2018                  & 3                        & 0.884  & 0.796  & 0.778 & 5.47       & 6.88       & 2.94       \\
			\cite{RN96} & 2019                  & 1                        & 0.888  & 0.837  & 0.833 & 4.62       & 4.13       & 2.65       \\
			\cite{RN97} & 2019                  & 2                        & 0.883  & 0.861  & 0.810 & 4.80       & 4.21       & 2.45       \\
			\cite{RN98} & 2019                  & 3                        & 0.890  & 0.830  & 0.810 & 4.85       & 3.99       & 2.74       \\ \hline
		\end{tabular}
	
	N/A: not available, WT: Whole tumor, TC: Tumore core, and ET: Enhancing tumor.
	\end{table}
	
	In the past three years, BraTS has also focused on prediction of OS. Table 5 lists the top three results for the OS prediction task.  RF regression was a popular method for this task. Shboul et al. \cite{RN99} extracted 1366 textures and other features, selecting significant features in three steps. The 40 most significant features were used to train the RF regression model and predict OS.  Puybareau et al. \cite{RN103} extracted features from segmented tumor region and introduced patient age into the feature space. PCA was performed to normalize the training set. The feature-wise mean, standard deviation, and projection matrix (W) were computed and stored during the rescaling phase of the PCA. The RF regression model was trained based on the normalized data. The feature vector of the test set was also normalized by the feature-wise mean and standard deviation derived from the training phase, and was then projected in the principal component space with W. The rescaled vectors were fed into the trained RF classifiers and the final prediction was obtained by majority voting. Sun et al. \cite{RN104} extracted 4526 features from the tumor lesion based on their previous segmentation results. Important features were selected by decision tree and cross-validation. Finally, they trained an RF regression model to predict OS. MLP was another popular method for this task. Jungo et al. \cite{RN100} computed 26 geometrical features from the segmented tumor regions and added age to complete the feature space. The four most important features were selected before being fed into a fully-connected neural network with one hidden layer and a linear activation function. Baid et al. \cite{RN105} extracted features from segmented tumor regions and excluded high-correlation features by Spearman correlation. An MLP was trained with variables that demonstrated statistically significant correlation with OS. He et al. \cite{RN110} selected seven features as input for a fully-connected neural network with two hidden layers. Their linear regression model also achieved good results. Feng et al. \cite{RN102} extracted image features and non-imaging clinical features to construct a linear regression model. They used two-dimensional feature vectors to represent the non-image features of resection status and compensate for sparse resection status data. They used a linear regression model to fit the training data after feature normalization. Weninger et al. \cite{RN106}  measured the volume of subregions based on segmentation results. The volume information, the distance between the centroids of tumor and brain, and patient age were used as input for linear regression to predict OS.  In addition to radiomic features, Wang et al. \cite{RN108} also considered biophysical modeling of tumour growth and calculated the ratio of second semi-axis length between TC and WT, to define a novel measure termed the relative invasiveness coefficient (RIC). Following feature selection, RIC, age and radomic features were fed into the epsilon-support vector regression. The method achieved an accuracy of 0.56 in OS prediction by incorporating RIC. 
	
	\begin{table}[]
		\caption{Overview of the studies and results with top 3 OS prediction perfromance of each year from 2017 to 2019.}
		\begin{tabular}{lllllll}
			\hline
			Ref                           & Year & Ranking & Accuracy & MSE      & Median-SE & Std-SE    \\ \hline
			\cite{RN99}  & 2017 & 1       & 0.579    & 245779.5 & 24944.4   & 726624.7  \\
			\cite{RN100} & 2017 & 2       & 0.568    & 213000.0 & 28100.0   & 662600.0  \\
			\cite{RN101} & 2017 & 3       & N/A      & N/A      & N/A       & N/A       \\
			\cite{RN102} & 2018 & 1       & 0.612    & 231746.0 & 34306.4   & N/A       \\
			\cite{RN103} & 2018 & 2       & 0.605    & N/A      & N/A       & N/A       \\
			\cite{RN104} & 2018 & 2       & 0.605    & N/A      & 32895.1   & N/A       \\
			\cite{RN105} & 2018 & 3       & 0.558    & 338219.4 & 38408.2   & 939986.8  \\
			\cite{RN106} & 2018 & 3       & 0.558    & 277890.0 & 43264     & N/A       \\
			\cite{RN107} & 2019 & 1       & 0.579    & 374998.8 & 46483.36  & 1160428.9 \\
			\cite{RN108} & 2019 & 2       & 0.56     & N/A      & N/A       & N/A       \\
			\cite{RN109} & 2019 & 3       & 0.551    & N/A      & N/A       & N/A       \\
			\cite{RN110} & 2019 & 3       & 0.551    & 41000.0  & 49300.0   & 123000.0  \\ \hline
		\end{tabular}
	
	N/A: not available. MSE: Mean square error
	\end{table}

	\noindent 
	\subsection{Unsupervised learning in tumor subregion analysis of medical images}
	
	Unsupervised learning has also been widely used in tumor subregion analysis of medical images for data without available or well-defined labelled training dataset. The twenty-five papers employing unsupervised learning techniques listed in Table 6 most focus on OS and PFS prediction and identification of tumor recurrence. There are several widely-adopted unsupervised algorithms, including level set methods (LSM), thresholding, individual- and population-level clustering, and K-means.  
	
	\noindent 
	\subsubsection{Level Set Methods}
	
	Level set methods are commonly used for unsupervised learning applied to segmentation tasks. Cui et al. \cite{RN111} developed and validated prognostic imaging biomarkers to predict OS of GBM patients based on multi-region quantitative image analysis. Each tumor was semi-automatically delineated by the level set algorithm and the segmented lesion was further divided into several subregions based on the hidden Markov random field (MRF) model and the EM algorithm \cite{RN33}. The biomarker was generated based on LASSO to predict the OS of the patients with GBM, and the model was tested by an independent cohort from the local institution. The concordance index and stratification of OS using the log-rank test were 0.78 and P = 0.018  for the proposed method, outperforming conventional prognostic biomarkers such as age (concordance index: 0.57, P = 0.389) and tumor volume (concordance index: 0.59, P = 0.409). In a later study, Cui et al. \cite{RN113} defined a high-risk volume (HRV) based on mpMRI images for predicting GBM survival and investigated its relationship and synergy with molecular characteristics. Each tumor was delineated by the level set algorithm and manual correction was performed for eight failed cases. The patients with an unmethylated MGMT promoter and high HRV had significantly shorter OS (median 9.3 vs. 18.4 months, log-rank P = 0.002), indicating the volume of the high-risk intratumoral subregion identified on mpMRI can predict survival and complement genomic information.
	
	\noindent 
	\subsubsection{Threshold-based Methods}
	
	Threshold algorithms are also suitable to separate tumor subregions based on imaging characteristics. Lawrence et al. \cite{RN114} investigated whether three month treatment response of newly diagnosed GBM based on C-methionine-positron emission tomography (MET-PET) could better predict prognosis than baseline MET-PET or anatomic magnetic resonance imaging alone. A threshold of 1.5 times mean cerebellar uptake was used to automatically segment the metabolic tumor volume (MTV). Persistent MTV at three months was defined as the overlap of the three month MTV and the pre-treatment MTV. Cox proportional hazards was used to perform multivariate analysis of PFS and OS. The results showed that most patients (67\%) with gross total resection (GTR) of newly diagnosed GBM  have measurable postoperative MTV and that the total and persistent MTV three months post-CRT were predictors of PFS. GTV-Gd at recurrence encompassed 97\% of the persistent MET-PET subvolume, 71\% of the baseline MTV, 54\% of the baseline GTV-Gd, and 78\% of the three month MTV, respectively. The persisitent MET-PET subvolume best predicts the location of tumor recurrence.  Legot et al. \cite{RN115} developed a framework to identify the tumor subregions of head and neck squamous cell carcinoma (HNSCC) with the risk of high recurrence on 18F-FDG PET images so that these might be considered for CRT dose escalation. Follow-up 18F-FDG PET images were registered with baseline images using an automatic rigid registration algorithm based on mutual information. Seven metabolic tumor regions were segmented in baseline images by characteristic fixed percentages of ${SUV}_{max}$ and compared with two post-treatment subregions of local recurrence or residual metabolic activity. The overlap between metabolic tumor subregions derived from baseline and follow-up PET images was only moderate.
	
	Estrogen Receptor (ER) status is a recognized molecular feature of breast cancer correlated with prognosis and its early detection can significantly improve treatment efficacy by guiding selection of targeted therapies \cite{RN116}. Chaudhury et al. developed a novel framework to classify ER status by extracting textural kinetic features from peripheral and core tumor subregions \cite{RN117} The WT was segmented using automatic threshold selection \cite{RN118} combined with morphological dilation and connected component analysis. The WT was divided into two subregions according to tumor geometry. Two feature selection methods (wrapper \cite{RN119} and correlation-based feature subset selection (CFS) \cite{RN120}) and three classifiers (naive Bayes  \cite{RN121}, SVM  \cite{RN29, RN122}, decision tree \cite{RN123}) were adopted in this study and each feature selector followed a classifier, for a total of six model composition combinations. The best classification accuracy approached 94\%, indicating that sub-region texture feature extraction can accurately classify ER status.
	
	\noindent 
	\subsubsection{Individual- and population-level clustering}
	
	Individual- and population-level clustering are used to assign each pixel or voxel to suitable clusters in order to divide a tumor into subregions. After tumor subregions are obtained, the relationship between tumor subregions, OS and PFS can be investigated.
	
	Wu et al. used individual- and population-level clustering in three works related to tumor subregion analysis. In one of their studies \cite{RN124}, they developed a robust tumor partitioning method to identify clinically relevant, high-risk subregions in lung cancer. The method divided the tumor into subregions based on a two stage clustering process: it first performed patient-level over-segmentation of the tumor into superpixels via K-means clustering \cite{RN36} on both PET and CT images, then these superpixels were merged to subregions via population-level hierarchical clustering \cite{RN126}. High-risk subregions predicted OS and out-of-field progression (OFP) over the entire cohort with a C-index of 0.66-0.67. For patients with stage III disease, the C-index reached 0.75 (HR 3.93, log-rank P < 0.002) and 0.76 (HR 4.84, log-rank P < 0.002) for predicting OS and OFP, respectively. In contrast, the C-index was lower than 0.60 for traditional imaging markers. The results showed that the volume of the most metabolically active and heterogeneous solid components of the tumor could predict OS and OFP better than conventional imaging markers. In a second study, Wu et al. \cite{RN127} developed an imaging biomarker to assess early treatment response and predicted outcomes in oropharyngeal squamous cell carcinoma (OPSCC). Based on 18F-FDG PET and contrast CT imaging, the primary tumor and involved lymph nodes were divided into subregions by individual- and population-level clustering. The proposed imaging biomarker was generated by the LASSO algorithm. The C-index was 0.72 for the training set and 0.66 for the validation set, suggesting the proposed biomarker can accurately predict disease progression and provide patients with better risk-adapted treatment. In a third study investigating risk-stratification in breast cancer, Wu et al. divided each tumor into multiple spatially segregated, phenotypically consistent subregions based on individual- and population-level clustering, and used a net strategy to construct an imaging biomarker based on image features derived from the multiregional spatial interaction (MSI) matrix \cite{RN128}. The results showed that breast cancers may exhibit three intratumoral subregions with distinct perfusion characteristics, and tumor heterogeneity may be an independent predictor of recurrence-free survival (RFS), independent of traditional predictors.
	
	In order to predict PFS in patients with nasopharyngeal carcinoma (NPC), Xu et al. extracted subregion features via individual- and population-level clustering to generate a biomarker by LASSO \cite{RN129}. Three subregions ($S_1$, $S_2$, $S_3$) with distinct PET/CT imaging characteristics were obtained. The C-index and log-rank test for imaging biomarker  $S_3$ and WT are 0.69 and 0.58, and  P < 0.001 and P < 0.552, respectively, indicating $S_3$ is superior to WT in terms of prognostic performance. Imaging biomarker $S_3$ and American Joint Committee on Cancer (AJCC) stages III–IV were identified as independent predictors of PFS based on multivariate analysis (P=0.011 and P=0.042, respectively). When combined to form a scoring system, imaging biomarker $S_3$ and AJCC stages III-IV outperformed AJCC staging alone (log-rank test P < 0.0001 vs. 0.0002;  P < 0.0021 vs. 0.0277 for the primary and validation cohorts, respectively). The results demonstrated that PET/CT subregion radiomics was able to predict PFS in NPC and provide prognostic information to complement other established predictors.
	
	Even et al. \cite{RN130} designed a subregional analysis for non-small cell lung cancer (NSCLC) using multi-parametric imaging. The multi-parametric images were divided into subregions in two clustering steps: each tumor was first divided into homogeneous subregions (i.e. super voxels) before being segregated into phenotypic groups by hybrid hierarchical clustering \cite{RN37}. Patients were clustered according to the absolute or relative volume of super voxels. The results showed that hypoxia, FDG avidity, and an intermediate level of blood flow/blood volume indicated a high-risk tumor type with poorer survival (P=0.035), providing evidence of the prognostic utility of subregion classification based on multi-parametric imaging in NSCLC.
	
	\noindent 
	\subsubsection{K-means}
	
	K-means is a popular unsupervised learning method that partitions samples into k clusters. Xie et al. developed a survival prediction model for patients with oesophageal squamous cell carcinoma (OSCC) prior to concurrent CRT \cite{RN132}. The patient’s tumor regions were divided into subregions by K-means clustering. Radiomic features were then extracted from these sub-regions to construct a biomarker based on the LASSO algorithm and predict OS. Independent patient cohorts from another hospital were used to validate the model. The C-indices were 0.729 (0.656–0.801, 95\% CI) and 0.705 (0.628–0.782, 95\% CI) in the training and validation cohorts, respectively. AUC values for the 3-year survival ROC were 0.811 (0.670–0.952 95\% CI) and 0.805 (0.638–0.973, 95\% CI), respectively. Such a model may facilitate personalized treatment through accurate prediction of early treatment response.
	
	Torheim et al. used K-means in MRI imaging of cervical cancer to divide voxels into two clusters based on relative signal increase (RSI) time series. Clusters of hypo-enhancing voxels demonstrated a significant correlation with locoregional recurrence (P=0.048) \cite{RN133}. Tumors with poor treatment response exhibited this characteristic in several regions, indicating a potential candidate for targeted radiotherapy.
	
	Franklin et al. developed a method to semi-automatically segment viable and non-viable tumor regions in colorectal cancer based on DEC-MRI, and compared these with histological subregions of viable and non-viable tumor, analyzing extracted pharmacokinetic parameters between them \cite{RN134}. The WT was manually delineated and four sub-regions were automatically obtained by PCA, followed by K-means. These four subregions were manually merged into two: viable and non-viable tumor. For viable tumor subregions defined by imaging and histology, DSC = 0.738 indicating the consistency of viable tumor segmentation between pre-operative DCE-MRI and postoperative histology. This technique may facilitate non-invasive assessment of treatment response in clinical practice.
	
	\noindent 
	\subsubsection{Others}
	
	Seow et al.  \cite{RN135} segmented the solid subregion of high-grade gliomas in MRI images by active contour modeling (ACM). The different ratio $((s_{ACM}-S_{manual})/\ s_{ACM}$, where $s_{ACM}$ and $s_{manual}$ are segmented area of ACM and manual, respectively) is 1.3. This algorithm produced segmentations in under twenty minutes, while manual segmentation required an hour, demonstrating suitability for efficient segmentation of solid enhancing regions in the glioma tumor core.  
	Fan et al. developed a framework to assess intratumoral heterogeneity in breast cancer based on the decomposition of DCE-MR images \cite{RN138}. The whole breast tumor was segmented by the fuzzy C-means (FCM) algorithm \cite{RN139}. A convex analysis of mixtures (CAM) method was then used to differentiate heterogeneous regions. Imaging features extracted from these regions were used predict prognosis and identify gene signatures. The results showed that tumor heterogeneity was negatively correlated with survival and the presence of cancer-related genetic markers of breast cancer.
	Wang et al. studied primary and secondary intrahepatic malignancies to determine whether an increase in tumor subvolume with elevated arterial perfusion during RT can predict tumor progression following treatment \cite{RN140}. The arterial perfusion of tumors prior to treatment were clustered into low-normal and elevated perfusion by global-initiated regularized local fuzzy clustering (GIRLFC) \cite{RN161}. The tumor sub-volumes with elevated arterial perfusion were extracted from the hepatic arterial perfusion images. The changes in tumor sub-volumes and arterial perfusion averaged over the tumors from pre-treatment baseline to mid-treatment were investigated for prediction of tumor progression following treatment. The results showed that an increase in intrahepatic subvolume with elevated arterial perfusion during RT may be a predictor of post-treatment tumor progression (AUC = 0.9).
	Lucia et al. \cite{RN38} developed a framework to evaluate the overlap between the initial high-uptake sub-volume ($V_1$) on baseline 18F-FDG PET/CT images and the metabolic relapse ($V_2$) after chemoradiotherapy in locally advanced cervical cancer. CT images of recurrence were registered with baseline CT using the 3D Slicer Expert Automated Registration module \cite{RN3802} to obtain the deformation fields by optimizing the Mattes mutual information metric \cite{RN41}, and the corresponding PET images were registered using the corresponding deformation fields. The fuzzy locally adaptive Bayesian (FLAB) algorithm \cite{RN39} was used to determine the sub-volumes $V_1$ and $V_2$ for baseline and follow-up PET images. The overlaps between the baseline high-uptake sub-volume and the recurrent metabolic volume were moderate to good (range (mean ± std)): 0.62–0.81 (0.72 ± 0.05), 0.72–1.00 (0.85 ± 0.10), 0.55–1.00 (0.73 ± 0.16) and 0.50–1.00 (0.76 ± 0.12) for DSC, overlap fraction, X ($X=\frac{V_1\cap V_2}{V_1}$) and Y($Y=\frac{V_1\cap V_2}{V_2}$), respectively.
	
	\begin{table}[]
		\caption{Unsupervised learning for tumor subregions analysis. HN: head and neck.}
		\begin{tabular}{llllllll}
			Ref                                            & Year                  & Models                                                                                                              & Modality                                                                 & Task                                                                                              & ROI                    & \begin{tabular}[c]{@{}l@{}}\# of patients in \\ training/testing datasets\end{tabular} &  \\ \cline{1-7}
			\cite{RN111}                  & 2016                  & Level set, MRF, EM                                                                                                  & Post-T1W, FLAIR                                                          & Predict OS                                                                                        & Brain                  & 46/33                                                                                  &  \\
			\cite{RN113}                  & 2017                  & level set                                                                                                           & T1W-ce, DWI                                                              & Predict OS                                                                                        & Brain                  & 62/46                                                                                  &  \\
			\cite{RN114}                  & 2019                  & Threshold, Cox proportional hazards                                                                                  & (11)C-MET-PET, T1W-Gd, FLAIR                                             & \begin{tabular}[c]{@{}l@{}}Recurrence tumor identification, \\ predict PFS\end{tabular}           & Brain                  & 37 (N/A)                                                                               &  \\
			\cite{RN115}                  & 2018                  & Threshold                                                                                                           & 18F-FDG PET/CT                                                           & Recurrence volume identification                                                                  & HN                     & 38 (N/A)                                                                               &  \\
			\multirow{2}{*}{\cite{RN117}} & \multirow{2}{*}{2014} & \multirow{2}{*}{\begin{tabular}[c]{@{}l@{}}Threshold, SVM, Nave Bayes, \\ decision tree, wrapper, CFS\end{tabular}} & \multirow{2}{*}{DCE-MRI}                                                 & \multirow{2}{*}{\begin{tabular}[c]{@{}l@{}}Estrogen receptor \\ (ER) classification\end{tabular}} & \multirow{2}{*}{Chest} & \multirow{2}{*}{20, LOOCV}                                                             &  \\
			&                       &                                                                                                                     &                                                                          &                                                                                                   &                        &                                                                                        &  \\
			\cite{RN124}                  & 2016                  & \begin{tabular}[c]{@{}l@{}}Individual- and population-level\\ clustering\end{tabular}                               & 18F-FDG /CT                                                              & Predict OS and OFD                                                                                & chest                  & 44 (N/A)                                                                               &  \\
			\cite{RN127}                  & 2020                  & \begin{tabular}[c]{@{}l@{}}Individual- and population-level\\ clustering\end{tabular}                               & 18F-FDG PET/CT                                                           & \begin{tabular}[c]{@{}l@{}}Assess early response \\ and predict PFS\end{tabular}                  & HN                     & 162, 10 fold                                                                           &  \\
			\cite{RN128}                  & 2018                  & \begin{tabular}[c]{@{}l@{}}Individual- and population-level\\ clustering\end{tabular}                               & DCE-MRI                                                                  & Predict RFS                                                                                       & Chest                  & 60/186                                                                                 &  \\
			\multirow{2}{*}{\cite{RN129}} & \multirow{2}{*}{2019} & \multirow{2}{*}{\begin{tabular}[c]{@{}l@{}}Individual- and population-level \\ clustering LASSO\end{tabular}}       & \multirow{2}{*}{18F-FDG PET/CT}                                          & \multirow{2}{*}{Predict PFS}                                                                      & \multirow{2}{*}{HN}    & \multirow{2}{*}{85/43}                                                                 &  \\
			&                       &                                                                                                                     &                                                                          &                                                                                                   &                        &                                                                                        &  \\
			\cite{RN130}                  & 2017                  & \begin{tabular}[c]{@{}l@{}}Individual- and population-level\\ clustering\end{tabular}                               & \begin{tabular}[c]{@{}l@{}}PDG PET, CT, \\ DCE-MRI, HX4 PET\end{tabular} & Predict OS                                                                                        & Chest                  & 36 (N/A)                                                                               &  \\
			\cite{RN132}                  & 2019                  & K-means, LASSO                                                                                                      & CT                                                                       & Predict OS                                                                                        & HN                     & 87/46                                                                                  &  \\
			\cite{RN133}                  & 2016                  & k-means                                                                                                             & DCE-MRI                                                                  & Recurrence tumor identification                                                                   & Pelvis                 & 81 (N/A)                                                                               &  \\
			\cite{RN134}                  & 2020                  & K-means, PCA                                                                                                        & DCE-MRI                                                                  & \begin{tabular}[c]{@{}l@{}}Tumor subregion\\ segmentation\end{tabular}                            & Abdoen                 & 14 (N/A)                                                                               &  \\
			\cite{RN135}                  & 2016                  & ACM                                                                                                                 & \begin{tabular}[c]{@{}l@{}}Post-T1W, FLAIR,\\ T2W\end{tabular}           & \begin{tabular}[c]{@{}l@{}}Tumor subregion\\ segmentation\end{tabular}                            & Brain                  & 4 (N/A)                                                                                &  \\
			\cite{RN17}                   & 2018                  & K-means                                                                                                             & DCE-MRI                                                                  & Predict prognois                                                                                  & Chest                  & 77, LOCCV                                                                              &  \\
			\multirow{2}{*}{\cite{RN138}} & \multirow{2}{*}{2019} & \multirow{2}{*}{FCM, CAM}                                                                                           & \multirow{2}{*}{DCE-MRI}                                                 & \multirow{2}{*}{Predict OS and RFS}                                                               & \multirow{2}{*}{Chest} & \multirow{2}{*}{61/173/87}                                                             &  \\
			&                       &                                                                                                                     &                                                                          &                                                                                                   &                        &                                                                                        &  \\
			\cite{RN140}                  & 2014                  & GIRLFC                                                                                                              & DCE-MRI                                                                  & \begin{tabular}[c]{@{}l@{}}Predict tumor progression\\ after RT\end{tabular}                      & Abdoen                 & 20 (N/A)                                                                               &  \\
			\cite{RN160}                  & 2019                  & FLAB                                                                                                                & 18F-FDG PET/CT                                                           & \begin{tabular}[c]{@{}l@{}}Tumor subregion\\ segmentation\end{tabular}                            & HN                     & 54 (N/A)                                                                               &  \\
			\cite{RN161}                  & 2012                  & GIRLFC                                                                                                              & DCE-MRI                                                                  & \begin{tabular}[c]{@{}l@{}}Predict subvolume related to\\ treatment outcome\end{tabular}          & HN                     & 14 (N/A)                                                                               &  \\
			\cite{RN162}                  & 2019                  & 3D Level set                                                                                                        & 18F-FDG PET/CT                                                           & Predict OS                                                                                        & Chest                  & 30 (N/A)                                                                               &  \\
			\cite{RN163}                  & 2018                  & PCA                                                                                                                 & \begin{tabular}[c]{@{}l@{}}DCE-MRI, DWI,\\ PET/CT\end{tabular}           & \begin{tabular}[c]{@{}l@{}}Predict neoadjuvant\\ therapy response\end{tabular}                    & Chest                  & 35 (N/A)                                                                               &  \\
			\cite{RN164}                  & 2019                  & CAM, RF                                                                                                             & DCE-MRI                                                                  & \begin{tabular}[c]{@{}l@{}}Predict breast cancer\\ subtypes\end{tabular}                          & Chest                  & 211. LOOCV                                                                             &  \\
			\cite{RN165}                  & 2020                  & TTP, SVM, LASSO                                                                                                     & DCE-MRI                                                                  & \begin{tabular}[c]{@{}l@{}}Predict HER2 2+ statuss\\ in breast cancer\end{tabular}                & Chest                  & 76, LOOCV                                                                              &  \\
			\cite{RN38}                   & 2020                  & FLAB                                                                                                                & 18F-FDG PET/CT                                                           & \begin{tabular}[c]{@{}l@{}}Recurrence tumor\\    \\ identification\end{tabular}                   & Plevis                 & 21 (N/A)                                                                               &  \\
			\cite{RN226}                  & 2019                  & K-means                                                                                                             & DWI, PET                                                                 & Segmentation and Predict PFS                                                                      & Chest                  & 18, LOOCV                                                                              &  \\ \cline{1-7}
		\end{tabular}
	\end{table}

	\noindent 
	\section{PREVALENCE OF METHODS}
	
	We have analyzed the percentage distribution of some attributes including the region of interest (ROI), learning strategy (supervised/unsupervised), technique (deep learning/non-deep learning), and imaging modalities (single/multi) (Figure 2). Brain and chest sites are the most studied regions of interest, with brain being most studied overall, likely in part due to the BraTS challenge providing public data as well as ground-truth for the non-public data.  Supervised learning accounts for 72\% of works reviewed, owing to the greater reliability and transparency of training when groud truth is available. The category of multi-modal studies account for 85\% of all works while the single-modality accounts for 15\%. A Non deep-learning strategy is employed in 61\%  of the summarized studies.
	
	\begin{figure}
		\centering
		\noindent \includegraphics*[width=6.50in, height=4.20in, keepaspectratio=true]{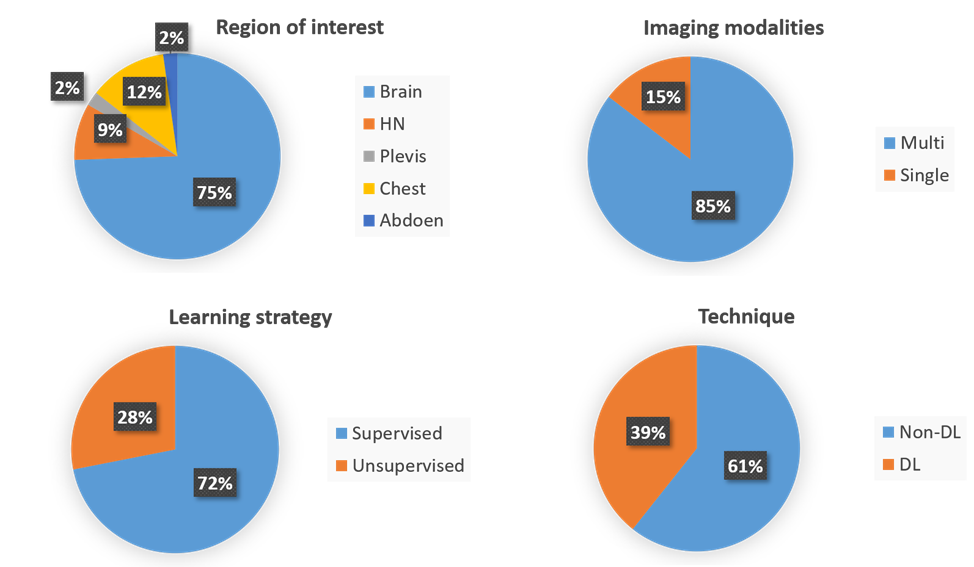}
		
		\noindent Figure 2. Pie charts for the distribution of various methods in AI-based tumor subregion analysis in medical imaging. HN: head and neck, DL: deep learning.
	\end{figure}
	
	\noindent 
	\section{SUMMARY AND OUTLOOK}
	
	AI methods from the field of computer vision have been widely adopted to complete several tasks in tumor subregion analysis.  As reviewed in here, brain is the most commonly studied site followed by chest. Since the subregions of brain tumors are generally accompanied by ground-truth data, supervised learning methods are more commonly employed than unsupervised strategies. For other body sites, unsupervised methods are more popular due to the lack of ground-truth.
	
	Currently, there is no universal image acquisition protocol for any imaging modality in clinical practice for sub-region analysis. Images acquired from different sites and scanners may affect the performance of these models. In order to address this issue, the quantitative imaging biomarkers alliance (QIBA) \cite{RN142} and the quantitative imaging network (QIN) \cite{RN143} have been working to formalize a standard imaging protocol.
	
	Sample sizes in the reviewed studies were small to intermediate (median (range): 230 (4-626)). For supervised learning, a large training set is required to train a reliable model. A large validation set is also essential in rigorously evaluating the proposed methods. Except for the BraTS studies, most reviewed here used institutional data and may lack generalizability. Many studies on tumor sub-regions demonstrate correlations to survival, as well as treatment response and recurrence. To validate these findings, significant time must be invested in follow-up especially in diseases with low overall mortality. Validation may also be confounded by adjuvant treatment during the follow-up period, complicating the analysis of any relationships that are discovered.
	
	Deep learning has demonstrated clinical utility in many tasks in medical imaging. At the time of writing, tumor subregion analysis is primarily in use for brain tumor subregion segmentation, but is rarely used in non-segmentation tasks or in other body sites. Great potential remains for DL applications in tumor subregion analysis. First, a CNN might be used to automatically extract useful features rather than relying upon handcrafted features. Secondly, for clinical tasks for which it is difficult to obtain manually-annotated ground truth data, an unsupervised CNN has been applied to solve the segmentation problem. As an example, Zhou et al. proposed a deep image clustering model to assign pixels to different clusters by updating cluster associations and cluster centers iteratively \cite{RN190}. Thirdly, CNN could be used to generate radiomic signatures for various clinical applications based on tumor subregion such as OS prediction, treatment response prediction and clinical risk stratification. In order to realize the full potential of DL applications in tumor subregion analysis, models must be trained on large datasets with external cross-site validation.

	\noindent
	\section{Summary and Discussion}
	
	GANs have been increasingly used in the application of medical/biomedical imaging. As reviewed in this chapter, cGAN- and Cycle-GAN-based image synthesis is an emerging active research field with all these reviewed studies published within the last few years. With the development in both artificial intelligence and computing hardware, more GAN-based methods are expected to facilitate the clinical workflow with novel applications. Compared with conventional model-based methods, GAN-based methods are more generalized since the same network and architecture for a pair of image modalities can be applied to different pairs of image modalities with minimal adjustment. This allows easy extension of the applications using a similar methodology to a variety of imaging modalities for image synthesis. GAN-based methods generally outperform conventional methods in generating more realistic synthetic images with higher similarity to real images and better quantitative metrics. In implementation, depending on the hardware, training a GAN-based model usually takes several hours to days. However, once the model is trained, it can be applied to new patients to generate synthetic images within a few seconds or minutes. Due to these advantages, GAN-based methods have attracted great research and clinical interest in medical imaging and biomedical imaging.
	
	Although the reviewed literatures show the success of GAN-based image synthesis in various applications, there are still some open questions that need to be answered in future studies. Firstly, for the training of GAN-based model, most of the reviewed studies require paired datasets, i.e., the source image and target image need to have pixel-to-pixel correspondence. This requirement poses difficulties in collecting sufficient eligible datasets, as well as demands high accuracy in image registration. As compared to cGAN, it is demonstrated that Cycle-GAN can relax the requirement of the paired datasets to be unpaired datasets, which can be beneficial for clinical application in enrolling large number of patient datasets for training. However, even the image quality derived by Cycle-GAN can be better than cGAN, the numerical performance may not be improved significantly in some synthesis tasks due to the residual mismatch between synthetic image and ground truth target image.
	
	Secondly, although the merits of GAN-based methods have been demonstrated, its performance can be inconsistent under the circumstances that the input images are drastically different from its training datasets. As a matter of fact, unusual cases are generally excluded in most of the reviewed studies. Therefore, these unusual cases, which do happen occasionally in clinic setting, should be dealt with caution when using GAN-based methods to generate synthetic image. For example, some patients have hip prosthesis. The hip prosthesis creates severe artifacts on both CT and MR images. The related effect of its inclusion in training or testing dataset towards network performance is an important question that has not been studied yet. There are more unusual cases that could exist in all those imaging modalities and are worth of investigation, just to name a few: all kinds of implants that introduce artifacts, obese patients whose scan has higher noise level on image than average, and patients with anatomical abnormality. To conclude, the research in image synthesis is still wide open. The authors are expected to see more activities in this domain for the years to come.

	\noindent 
	\bigbreak
	{\bf ACKNOWLEDGEMENTS}
	
	This research was supported in part by the National Cancer Institute of the National Institutes of Health under Award Number R01CA215718 and Emory Winship Cancer Institute pilot grant. 
	
	\noindent 
	\bigbreak
	{\bf CONFLICT OF INTEREST}
	
	The authors declare no conflicts of interest.

	\noindent 
	
	\bibliographystyle{plainnat}  
	\bibliography{arxiv}      
	
\end{document}